# Module control of network analysis in psychopathology

Chunyu Pan[1,2†], Quan Zhang[3,4†], Yue Zhu[1,5], Shengzhou Kong[1], Juan Liu[1,5], Changsheng Zhang[2], Fei Wang[1,5,6*], Xizhe Zhang[7,8*]

1. Early Intervention Unit, Department of Psychiatry, The Affiliated Brain Hospital of Nanjing Medical University, Nanjing, Jiangsu, 210024, China;
2. Northeastern University, Shenyang, Liaoning, 110169, China;
3. Vanke School of Public Health, Tsinghua University, Beijing, 100084, China;
4. Institute for Healthy China, Tsinghua University, Beijing, 100084, China;
5. Functional Brain Imaging Institute of Nanjing Medical University, Nanjing, Jiangsu, 210024, China;
6. Department of Mental Health, School of Public Health, Nanjing Medical University, Nanjing, Jiangsu, 211166, China;
7. School of Biomedical Engineering and Informatics, Nanjing Medical University, Nanjing, Jiangsu, 210033, China;
8. Lead contact;

†Chunyu Pan and Quan Zhang contributed equally to this work;

*Correspondence: Xizhe Zhang: zhangxizhe@njmu.edu.cn; Fei Wang: fei.wang@yale.edu;

**Highlights**

- Introduces Module Control, a novel concept for exploring the control mechanisms between network modules.
- Introduces a novel metric for assessing control powers between modules within psychopathology networks.
- Non-emotional modules, such as sleep-related and stress-related modules, are the primary controlling modules in the symptom network.

**Summary**

The network approach to characterizing psychopathology departs from traditional latent categorical and dimensional approaches. Causal interplay among symptoms contributed to dynamic psychopathology system. Therefore, analyzing the symptom clusters is critical for understanding mental disorders. Furthermore, despite extensive research studying the topological features of symptom networks, the control relationships between symptoms remain largely unclear. Here, we present a novel systematizing concept, module control, to analyze the control principle of the symptom network at a module level. We introduce Module Control Network (MCN) to identify key modules that regulate the network's behavior. By applying our approach to a multivariate psychological dataset, we discover that non-emotional modules, such as sleep-related and stress-related modules, are the primary controlling modules in the symptom network. Our findings indicate that module control can expose central symptom cluster governing psychopathology network, offering novel insights into the underlying mechanisms of mental disorders and individualized approach to psychological interventions.

**Graphical Abstract**

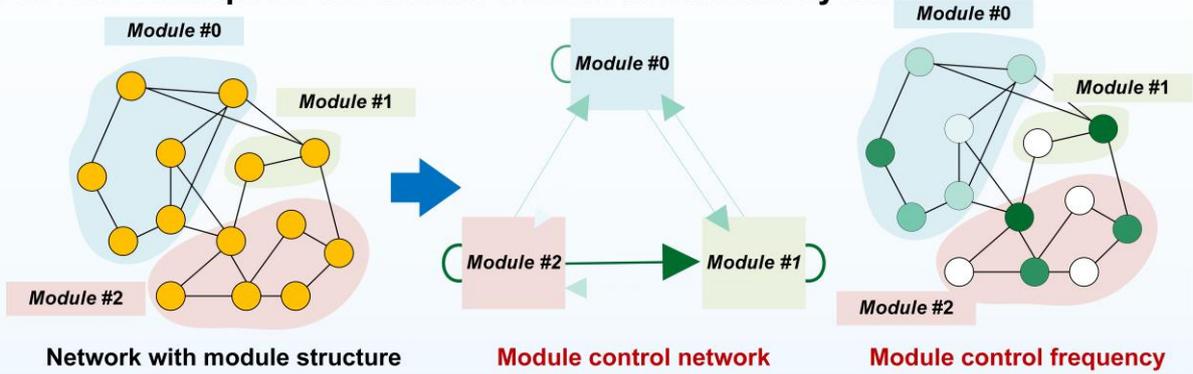
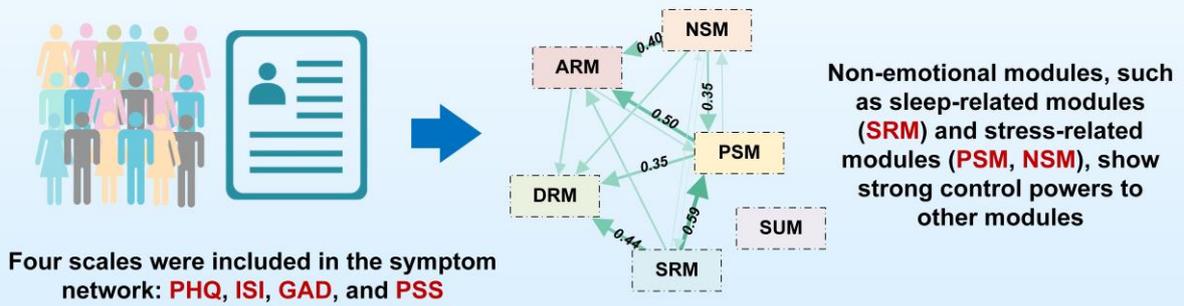
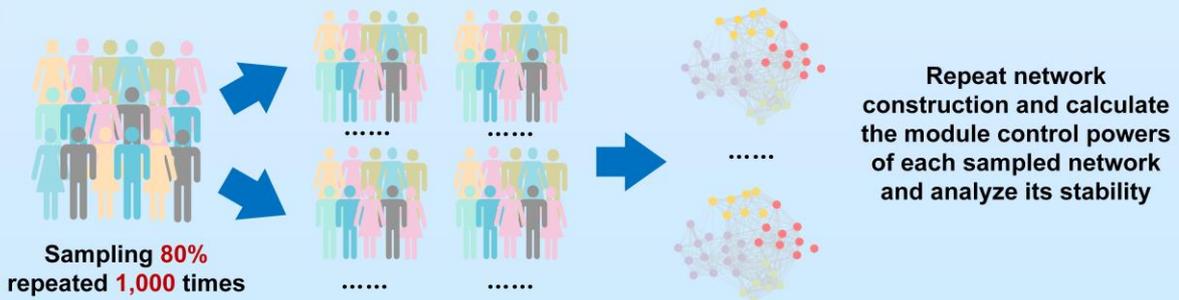

**Subject areas**

Symptom networks; module control; mental disorder; minimum dominating set;

## Introduction

The field of mental health research and intervention has experienced significant evolution, especially in how therapists and researchers approach symptoms and signs of mental disorders. Therapists often use subjective assessments such as the Patient Health Questionnaire (PHQ-9) and the Generalized Anxiety Disorder Scale (GAD-7) to understand a patient's symptoms[1, 2]. These tools aid in identifying the subjective symptoms experienced by patients. However, the primary challenge for both therapists and mental health researchers lies in accurately determining the underlying causes of a patient's signs and symptoms. In the broader history of psychiatry, monocausal and essentialist frameworks have long dominated psychiatric research. Neo-Kraepelinians described the signs and symptoms of psychopathology as objectively as possible to bolster the reliability of psychiatric diagnosis. Yet dissatisfaction with the Diagnostic and Statistical Manual of Mental Disorders (DSM) categorical system has grown over these years [3], and mental disorders has been defined as an increasingly labeled DSM diseases [4]. Complex mechanisms of co-morbidity among supposedly discrete conditions has become so troubling to clinicians that they have begun to look for new ways to explain what constitutes a mental disorder.

The network theory of psychopathology suggests that mental disorders are best understood as clusters of symptoms sufficiently unified by causal relations among those symptoms that support induction, explanation, prediction and control [5]. Signs and symptoms are constitutive of disorder, not the result of an unobservable common cause. In the past few decades, advancements in the field of psychopathology have yielded noteworthy strides through the application of network theory [6]. These research posits that symptoms of mental disorders do not exist in isolation, but instead, they form intricate systems of interconnected symptoms [6]. As a whole system, each symptom is not just mere manifestations of the mental disorder, but active contributors to its development and maintenance [7]. In other words, the symptoms constitute a mental disorder rather than being a product of a mental disorder. For now, this concept has been applied to a range of disorder, such as depression [8, 9], anxiety disorders [10], bipolar disorder [11], manic [11] and the general structure of psychiatric symptomatology [12].

However, traditional symptom network analysis focus more on the relationships among node or items of scales. This perspective overlooks the exploration of the relationships among symptom modules which are clustered by multiple symptoms. Module analysis within symptom networks can be employed to group items that are highly correlated into coherent subsets as clusters of symptoms. This approach allows for the investigation of potential interplay mechanisms among different disorders at a mesoscale level within mental illnesses. It provides valuable clues for further analyzing the functional aspects of symptom networks [13]. For instance, prior research conducted on the symptom network of major depressive disorder has delineated two discernible modules [14]. One module encompasses the core symptoms intricately tied to mood depression, including manifestations of profound sadness and anhedonia. The other module comprises symptoms that exhibit intricate interconnections with anxiety and disrupted sleep patterns, which evidently represent the core symptoms of comorbid conditions such as depression with anxiety disorders or insomnia. This study suggests that the specific multi-modularity of depression may also be a major reason for the varying treatment efficacy.

Furthermore, previous works in symptom networks analysis have primarily focused on the nodes with high centrality measures, such as degree centrality. However, for some networks, nodes of lower centrality might be more important to understand system behaviour [15]. This evolution in the paradigm of psychopathology studies, transitioning from an emphasis on individual symptoms to an analysis of symptom networks as dynamic [13, 16, 17], aim to understand the intricate interplay of symptoms within networks. Traditional centrality-based approaches are inadequate for unveiling the dynamic mechanisms within networks, highlighting the need for advanced control strategies. These strategies are essential to comprehend the networks' complexity and dynamism, particularly at a mesoscale level, such as the module-scale within symptom networks. Network controllability theory, especially structural controllability [18-22], emerges as a powerful approach in addressing these challenges. It takes the network system as a whole, makes inherent assumptions about the network system's operating mechanisms and looks for the critical nodes that driving the network state [18]. To date, network controllability theory has been applied in various biological fields, protein networks [23-25], brain networks [26-29], cancer [30-32], precise medication [33] and drug repurposing [34, 35], it makes possible for researchers to control the behavior of large-scale networks.

Here, we introduce *Module Control*, an innovative network control analytical concept designed to analysing the control associations of symptom clusters in the symptom network. We construct a new mesoscale network based on the control relationship between modules in the original network, i.e., the Module Control Network (MCN). Based on the distribution and frequency of driver nodes in the network module, we also establish a set of metrics for measuring module control capacities.Subsequently, we applied the Module Control method to a large-scale dataset of psychological screening data collected from university students, constructing a symptom network and analyzing its features. This work introduces the network control theory into psychiatric network analysis, combining the module characteristics of the network, thereby capturing the important dynamics of real-world psychological systems. This approach goes beyond evaluating traditional central symptoms and opens up new possibilities for treatment strategies, allowing interventions to be tailored to the unique dynamics of different disease patients.

## Result

### Module Control of Complex Network

Many networks exhibit a distinct module nature, where nodes within the same module are highly interconnected [36]. To addressing the intricate nature of module control within complex networks, we proposed the concept of module control, aiming to decipher the overarching control mechanisms that span different network scales.

Central to our investigation is the concept of the Minimum Dominating Set (MDSet), a subset of nodes whose control area spans the entire network. By selecting the node in the dominating set as the driver node, the entire network can be driven [37]. The smallest size of dominating set of the network is called minimum dominating set (MDSet) whose control is sufficient to fully control the system's dynamics with the lowest control cost (Figure 1.C, see STAR methods).

However, the MDSet is often not unique and it is known to be NP-hard problem [37] (Figure 1.E), there are only some approximation algorithms are available to solve this problem [38] currently. In large-scale networks, due to the complexity of calculating all MDSets, usually calculate random a part of the MDSets instead of all of them as a replacement for the results. This strategy, although effective, is not perfect, and compared to the overall mean of the MDSets, the mean of sampling still has a certain degree of error. Especially in small-scale networks that are structurally sensitive, it is particularly necessary to accurately calculate all the MDSets. Therefore, we employed a brute-force search method to discover all MDSets. As the small-scale network we used, we did not incur significant computational costs and were able to obtain all precise MDSets (see STAR methods).

Building on the foundation of MDSet analysis, we introduced the module control network (MCN), a novel conceptual network that reimagines the network's modular architecture through the lens of control theory (Figure 2). In the MCN, nodes represent the original network's modules, while edge weights indicate the Average Module Control Strength (AMCS) between modules, calculated across all identified MDSets (Figure 2.E). This innovative approach illuminates the asymmetrical control relationships between modules, revealing the directional nature of control flow within the network (see STAR methods).

Furthermore, we introduce the concept of control frequency (CF) for each node within a module, serving as an indicator of node importance in the network's control architecture. This metric, alongside the Average Module Control Frequency (ACF), provides valuable insights into the control capacity of each module, enabling a deeper understanding of the main controlling symptoms and items within a symptom network constructed from real-world data (see STAR methods).

Our exploration into module control within complex networks not only expands the theoretical framework of network controllability theory but also opens up new pathways for practical applications in psychopathology and beyond. By shifting the focus to modular control, we offer a fresh perspective on the dynamics of complex systems, paving the way for innovative intervention strategies targeted at the modular level.

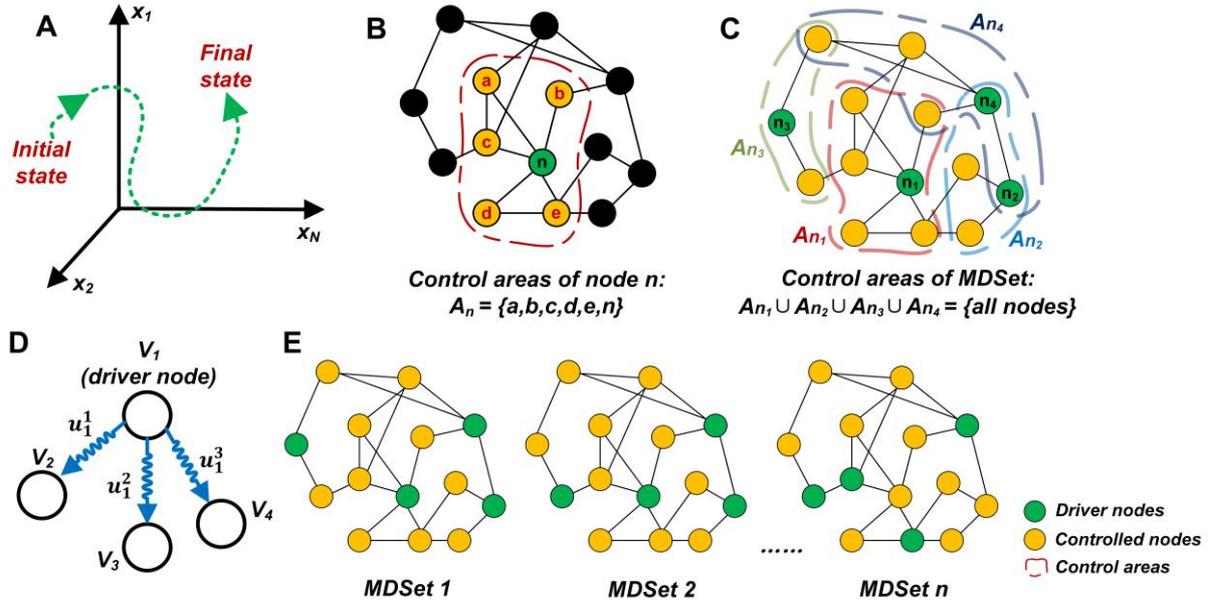

**Figure.1** Network controllability and minimum dominating set. **(A)** Illustration of network control. A network can be driven from any initial state to any final state in finite time. **(B)** Driver node and its control areas, which include its neighbor nodes and itself. **(C)** An MDSet of the network. It is a set with the minimum number of nodes and its union control areas can covering the entire network. **(D)** Dominating set control model. For an undirected network, a driver node can control all its neighbor nodes by different control signals, where $v_i$ and $u_j$ denote driver node and control signal, respectively. The blue arrow indicates the flow direction of the control signal, rather than the direction of the edges. **(E)** An example of different MDSets exist in the network.

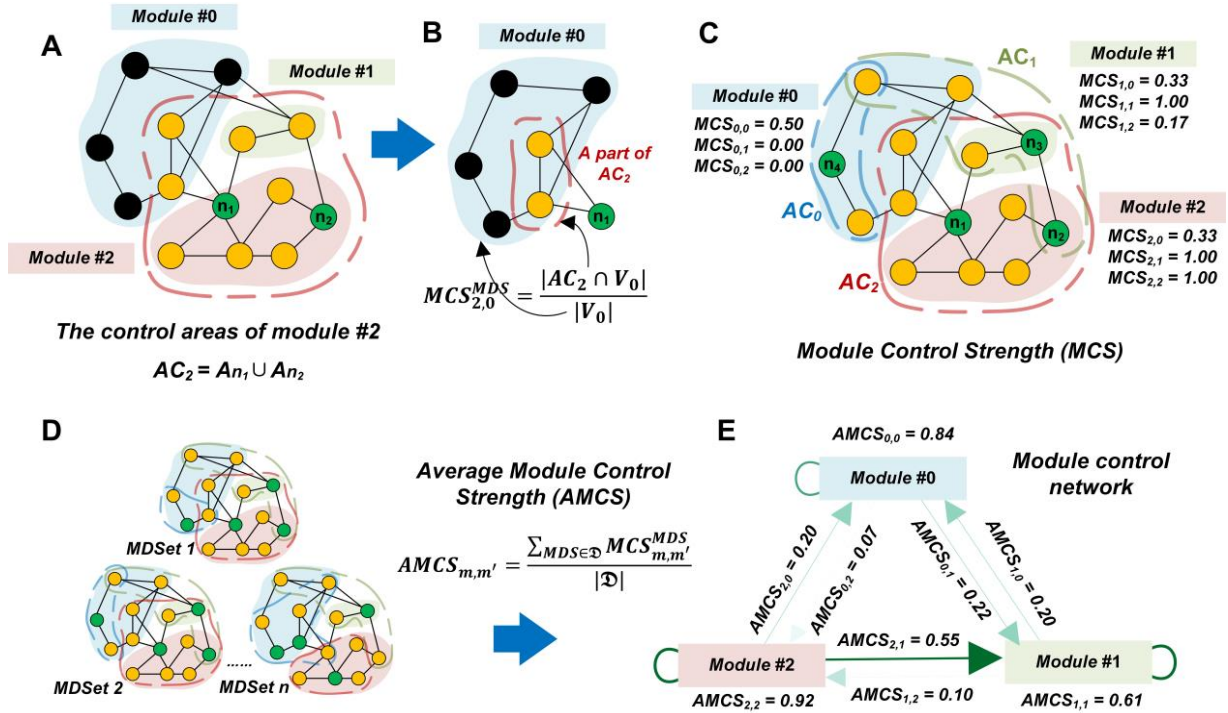

**Figure.2** An example of module control network. **(A)** An example of the control areas of module. The control areas of module $m$ is the union control areas of the driver nodes in module $m$. **(B)** Illustration of the module control strength (MCS). The MCS from module $m$ to $m'$ is the ability that the control areas of module m cover to module $m'$. **(C)** All MCS of a sample network. Any module pair will have an MCS and notice that $MCS_{m,m'}$ and $MCS_{m',m}$ are not equal due to the sizes of the modules and the distribution of the driver nodes. **(D)** A sample network with $n$ MDSets. Regardless of how the MDSet changes, the same network has fixed modules but has different driver nodes. **(E)** An example of module control network (MCN). The nodes in MCN are the modules in origin network and the edge weights represents the average module control strength (AMCS) based on $\mathcal{D}$ different MDSets.

**Topological Analysis of Symptom Networks**

In this study, we conducted symptom network using data from university students. A cohort of 2,773 college freshmen was recruited, consisting of 1311 (47.3%) males and 1462 (52.7%) females. Each participant was instructed to complete four self-assessment scales, encompassing the Generalized Anxiety Disorder-7 (GAD-7) [39], Insomnia Severity Index (ISI) [40], Patient Health Questionnaire-9 (PHQ-9) [41], and Perceived Stress Scale-14 (PSS-14) [42] (see STAR methods). These scales provided a comprehensive evaluation of the subjects' emotional states from four distinct perspectives, namely depression, anxiety, sleep, and stress. Consequently, these scales were utilized to construct the symptom network. The standard deviation (SD), mean, minimum (Min), maximum (Max), skewness and kurtosis of each item on the scales are provided in the supplementary material.

We construct the symptom network based on Graphical LASSO model [43] (see STAR methods), result in a network with 37 nodes and 254 edges (Figure 3.A). The edges in the network represent the inverse covariance of the two items in the precision matrix, with the average weight 0.258. The average degree of the network is 13.73, the network diameter is 3, the density of the network is 0.40 and the modularity is 0.27 (Figure 3.B). Upon performing z-score normalization based on network centrality, we identified several important symptoms based on network centrality (Figure 3.D). PSS3 is the node with the highest degree centrality (z=2.046), indicating its substantial influence on the network. Meanwhile, ISI1 exhibited the highest average strength (z=4.124), suggesting its significant connectivity with other nodes. PSS4 displayed the highest clustering coefficient (z=2.059), reflecting the presence of strongly interconnected structures in the network. Additionally, PSS7 showcased both the highest closeness centrality (z=1.302) and betweenness centrality (z=3.953), highlighting its importance as a connector and potential mediator between different network regions. The network's edge weights also revealed interesting patterns. The maximum edge weight of 1.568 occurred between ISI1 and ISI2, suggesting a strong relationship between these two symptoms. Conversely, the minimum edge weight of 0.0002 was observed between ISI2 and PSS14, indicating a weak association. Additional information about the edges and nodes can be found in the supplementary material.

Next, we conducted module identification for the symptom network, which led to the discovery of six distinct modules (Figure 3.A and 3.C). Module #0, associated with depression (abbreviated as **DRM**), encompasses seven items from the PHQ scale. Module #1, pertaining to sleep (abbreviated as **SRM**), includes eight items from the ISI scale and PHQ3 (sleep item). Module #2, exclusively connected to suicide (abbreviated as **SUM**), comprises a single item, PHQ9 (suicide item). Module #3, correlated with anxiety (abbreviated as **ARM**), incorporates all seven items from the GAD scale. Module #4, linked to negative stress (abbreviated as **NSM**), consists of eight items from the PSS scale, specifically PSS4, PSS5, PSS6, PSS7, PSS9, PSS10, PSS12, and PSS13. Of these items, PSS12 has a positive score, while the others possess negative scores. Lastly, Module #5, related to positive stress (abbreviated as **PSM**), comprises six items from the PSS scale, including PSS1, PSS2, PSS3, PSS8, PSS11, and PSS14, all of which exhibit positive scores.

Moreover, we assessed the significance of each module based on the average centrality value of nodes within the module, using six node properties, including degree, strength, closeness, k-core, clustering, and betweenness (Figure 3.E). Our analysis revealed that DRM possesses the highest k-core (z=0.649), while SRM exhibits the highest average strength (z=1.036). Since SUM contains only PHQ9, an item outside the primary components, it demonstrates the lowest value across all indicators. ARM shares the highest k-core with DRM (z=0.649) and also displays the highest clustering coefficient (z=0.891). PSM, on the other hand, features the highest degree centrality (z=0.940), closeness centrality (z=0.599), and betweenness centrality (z=1.223).

Lastly, we categorized the modules within the network into two distinct types based on their characteristics. Modules that do not directly describe emotional feelings, such as SRM, NSM, and PSM, are regarded as non-emotional modules. Conversely, modules explicitly describing emotional feelings, including DRM and ARM, are referred to as emotional modules. This classification provides a framework for understanding the diverse roles of modules within the symptom network.

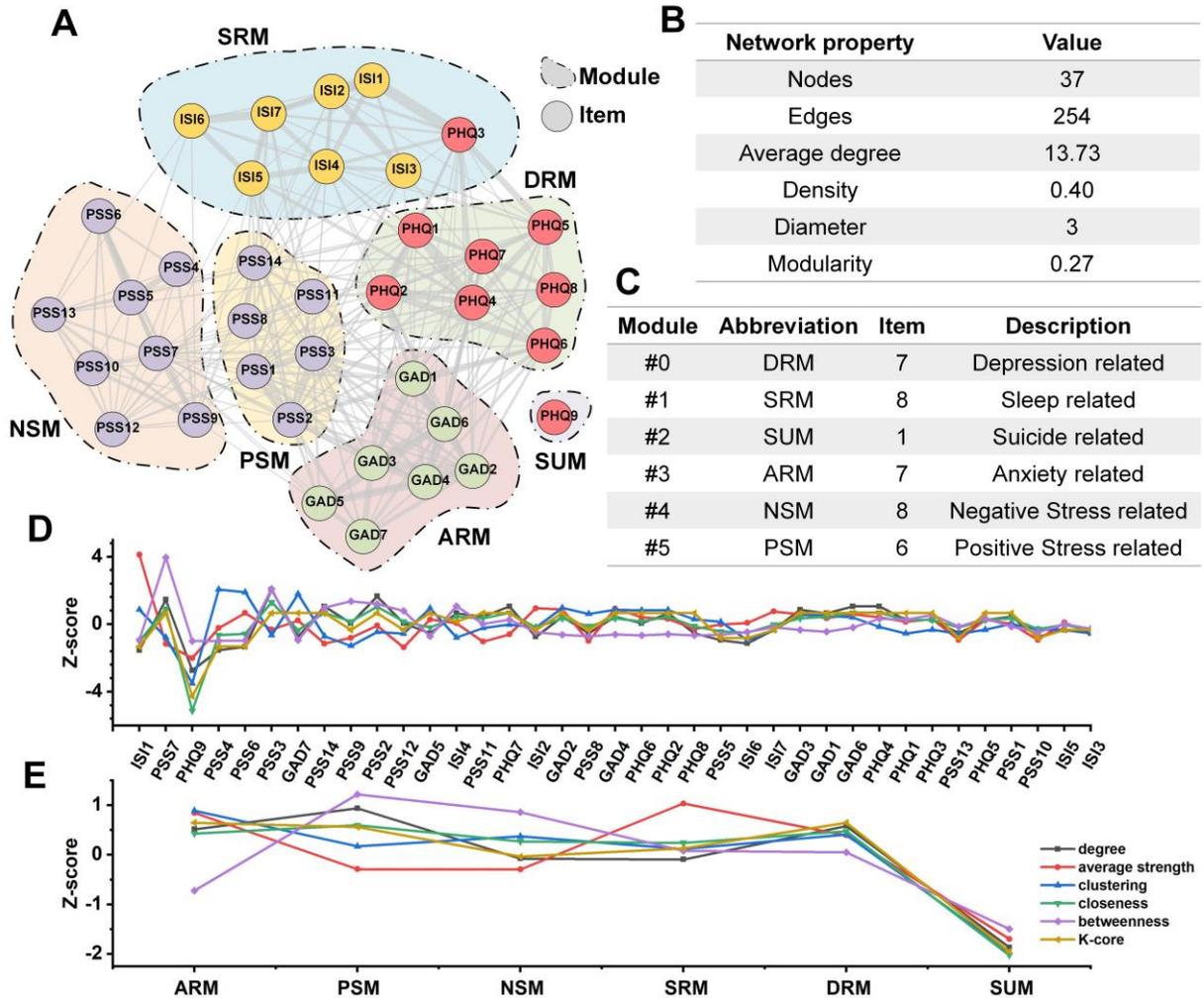

**Figure.3** The modules and properties of symptom network. (**A**) Network construction and modules detection results. The nodes in the network are marked with the same color if they belong to the same scale. Edges with larger weights are represented by larger size and greater color density. (**B**) Basic network properties. (**C**) Basic module properties. (**D**) Six network metrics for different nodes in the network. (**E**) Six metrics for different modules in the network. All detailed refer to the supplementary material.

**Module Control Analysis of Symptom Network**

In this section, we analyze the symptom network using module control, which is a network controllability analysis method based on Minimum Dominating nodes Set (MDSet). Although computing the MDSet is an NP-hard problem, considering the relatively small size of the symptom network, we were able to enumerate all of its MDSets (see STAR methods). In final, total of 27 MDSets were identified within the symptom network (supplementary material). We calculated the average weight of nodes in each MDSet and found that these MDSets exhibited similar weights (Mean=11.91, SD=1.134). Consequently, we computed the Average Module Control Strength (AMCS) for each module based on all MDSets and constructed the Module Control Network (MCN) in module scale (see STAR methods). The MCN consist 6 nodes and 36 edges, which the edge weight is the AMCS between modules (Figure 4.A).

Within the MCN, the three edges with the highest weights were identified as $AMCS_{SRM, PSM}$=0.586, $AMCS_{PSM, ARM}$=0.497, and $AMCS_{SRM, DRM}$=0.444. On the other hand, the three lowest $AMCS$ were $AMCS_{DRM, SRM}$=0.042, $AMCS_{ARM, NSM}$=0.042, and $AMCS_{DRM, NSM}$=0.019. Furthermore, our analysis revealed that the out-degree strength of non-emotional modules was higher than that of emotional modules in MCN, suggesting that non-emotional modules possess a greater capacity for controlling other modules (Figure 4.B). The in-degree strength of emotional modules was higher than that of non-emotional modules, indicating that emotional modules are more susceptible to controlled by other modules (Figure 4.C). These findings highlight the varying degrees of control exerted by different modules and emphasize potential connections and dependencies within the symptom network.

To further investigate these relationships, we analyzed four types of AMCS between non-emotional and emotional modules in MCN, encompassing the average AMCS of non-emotional modules to emotional modules, non-emotional modules to non-emotional modules, emotional modules to emotional modules, and emotional modules to non-emotional modules. Our analysis showed that the average AMCS from non-emotional modules to emotional modules was higher than the other types, whereas the average AMCS from emotional modules to non-emotional modules was the lowest (Figure 4.D). Collectively, these results indicate that the network generally exhibits a tendency for non-emotional modules to control emotional modules.

Furthermore, we calculated the control frequency (CF) of each node (Figure 4.E). Out of the 37 nodes, 24 nodes acted as driver nodes in one or more MDSets. The top five nodes with the highest CF are PHQ9 ($CF$=1.000), ISI4 ($CF$=0.778), PSS2 ($CF$=0.444), PSS7 ($CF$=0.370), and GAD6 ($CF$=0.222). Intriguingly, these top five nodes are situated within five distinct modules, indicating the core driver node of each module and ensuring the stability of control within each module. Furthermore, we computed the average control frequency (ACF) for each module, revealing that non-emotional modules exhibit a higher ACF than emotional modules (Figure 4.F). This suggests that non-emotional modules more frequently operate as driver nodes and assume a leading control role within the entire symptom network.

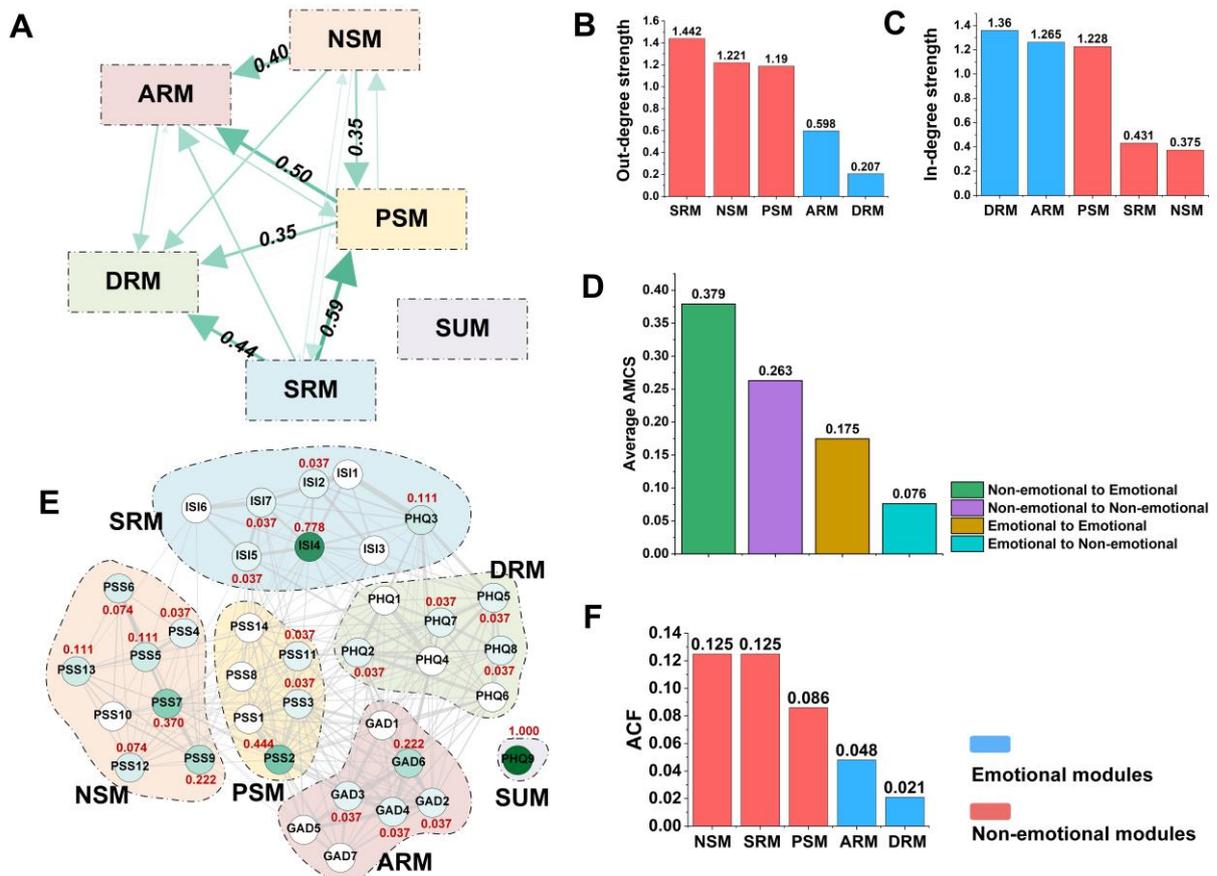

**Figure.4** The results of module control of symptom network. (**A**) The MCN of symptom network. The larger and bolder the arrow from any module, the greater the *AMCS* over the destination module. The top 5 *AMCS* are displayed. (**B**) The out-degree strength of MCN. For any module, the average *AMCS* is same as the module out-degree strength in MCN. (**C**) The in-degree strength of MCN. The average in-degree of each modules in MCN can be considered as the extent of it being controlled by other modules. Since we focus only on the module-module control capability, the out-degree and in-degree strengths

do not count the *AMCS* of each module self. **(D)** Average *AMCS* between non-emotional modules and emotional modules. **(E)** The control frequency of symptom network. **(F)** The *ACF* of different modules.

**The stability validation of module control**

The stability of the symptom network is crucial, as it has a significant impact on the results of control analysis. Consequently, we employed bootstrapping methods to validate the stability of the symptom network (see STAR methods). By randomly selecting 80% of the full sample, we constructed 1,000 network instances through bootstrapping. We then calculated the confidence intervals (CIs) for each edge. The results suggest that there was no significant difference in edge weights between the symptom network created using the full sample and the network generated by bootstrapping (Figure 5.A). Furthermore, the edge count in the network constructed with the full sample lies within the 95% confidence interval (Figure 5.B). In terms of node strength stability, the average correlation coefficient of node strengths between the bootstrapping networks and the full sample network remains highly stable, even as the number of sampled cases decreases (Figure 5.C). Remarkably, the average correlation of node strengths exceeds 0.8 in the network constructed with only 10% of the samples, indicating a high level of stability.

Considering that the size and number of MDSets and modules can significantly affect module control results, we calculated the number of MDSets and modules in the bootstrapping network. Our findings revealed that the number of MDSets in the symptom network constructed using the full sample falls within the 95% confidence interval (Figure 5.D), and the size of the MDSets in the bootstrapping network is approximately four (Figure 5.F), which is identical to that of the full sample network. Regarding the number of network modules, the majority of bootstrapping networks contained six modules (Figure 5.E), consistent with the results of the full sample network. The stability of the network modules and MDSets indicates that the control modes are consistent between both the bootstrapping and full sample networks.

We also utilized the average correlation coefficient based on case-dropping bootstrap to perform stability tests on the module control metrics, i.e. *AMCS* and *ACF*. The results indicate that the *AMCS* and *ACF* has a consistently stable average correlation coefficient even as the sample size decreases (Figure 5.C). The average correlation coefficient of *ACF* decreases and will lower than 0.5 when the sample size is less than 50% (see supplementary). This phenomenon is related to the long-tailed distribution of the number of MDSets (Figure 5.D). When the sample size used to construct the network is small, it exacerbates the uneven distribution of the number of MDSets and ultimately results in the instability of the average correlation coefficient.

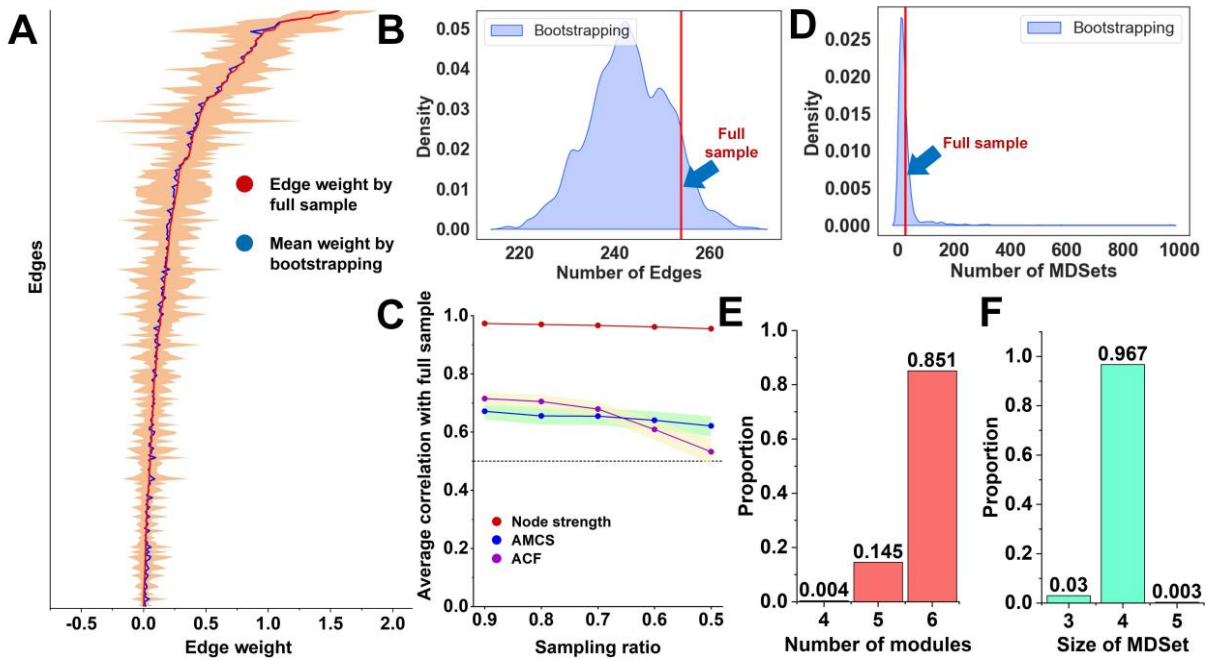

**Figure.5** Stability test of symptom network. **(A)** Full sample network value (red), bootstrapped 95% intervals (shaded area) and mean bootstrapping network value (blue) of edge weights. For detailed values of every edge, please refer to the supplementary material. **(B)** The edge number distribution of the bootstrapping network and full sample network. **(C)** Average correlation between full sample network and bootstrapping networks. Three different colors indicate different metrics. As an example, the red line is the results of node strength case-dropping bootstrap analysis, which shows the average correlation between strength centrality estimated in the full sample and strength estimated on a random subsample, retaining only a certain portion of cases (from 90% to 50%). Shaded area indicates 95% bootstrapped confidence intervals of correlation estimates. Higher values indicate better stability of centrality estimates. **(D)** The modules number distribution of the bootstrapping network and full sample network. **(E)** The module numbers distribution of the bootstrapping network. **(F)** The MDSet size distribution of the bootstrapping network.

## Discussion

Mental disorders are often characterized by a cluster of symptoms, with the underlying causes frequently remaining elusive. The causal interactions and interdependencies among these symptoms give rise to a complex network of interconnected symptoms and modularity phenomena. Modularity phenomena are prevalent in the majority of real-world networks, characterized by the presence of highly clustered nodes within distinct modules [44]. In this study, we employ an analytical approach rooted in network controllability theory to investigate the control capacity at the module scale within the symptom network. We construct a module control network (MCN), enabling us to understand and analyze control attributes between modules from a network scale perspective. We also propose a module control metrics, average module control frequency (*ACF*), to examine the control principles of symptom network. By applied module control metric to symptom networks among college students, we discovered that non-emotional modules, such as sleep and stress, play a leading control role in the entire network and may result in an individual's emotional dysregulation. This study provides the first evidence of control associations between different symptom modules. It lays the groundwork for understanding causal relationships within individual symptom networks, potentially aiding in the development of more effective and personalized interventions for mental disorders.

For symptom networks, extensive research has utilized network topology to identify central symptoms in the network [7]. Network topology metrics such as degree centrality, betweenness centrality and closeness centrality, derived from social networks, have been widely employed in various networks [45]. However, the application of these metrics to different networks may pose challenges. For instance, in estimation-based networks like symptom networks, the concept of "indirect correlation" or "path" can be unreliable. This is because the edges between nodes represent a correlation between node pairs, rather than a physical link. As the results, metrics based on the concept of "path" in symptom networks, such as betweenness and closeness, have been deemed unstable [6, 46, 47]. On the other hand, we believe that the direct relationship between a node and its neighbors in a symptom network is crucial. Metrics that assess the relationship between nodes and their direct neighbors, such as degree centrality and node strength, have been shown to be dependable [46]. The dominating set theory, which underpins module control, focuses precisely on the relationship between nodes and their direct neighbors in a network. The key concept of dominating set assumes that any node in the network can directly control its neighbors [37], without the need to delve into complex concepts like indirect relations or long paths. Therefore, the dominating set theory has been introduced as a fundamental technical tool for analyzing various biological correlation networks [25, 28, 48, 49].

The structural controllability model based on dominating sets offers several advantages, such as requiring an minimum nodes to driving network and more generalized assumptions about dynamics. Specifically, most treatments only focus on a subset of symptoms within the network currently, rather than addressing all of them. Therefore, it is crucial to effectively control the symptom network with as few control nodes as possible. Dominating set theory provides a way to identify the minimized number of driver nodes needed to control the network, making it to an effective method to control network. On the other hand, analyzing any real network as a complex system using network controllability methods requires making dynamics assumptions [18]. In the case of a network with linear dynamics, the network can be controlled through structural controllability proposed by *Liu* et al. [18]. However, because of the formidable size of real systems, a detailed complexity analysis of control dynamics in networks is impractical in most cases or limited to canonical linear time-invariant approximations at best [18]. In symptom networks, the complex dynamical processes of how symptoms interact with each other (such as depression and anxiety often accompany) remain unclear, making it challenging to assume specific dynamic processes for these networks. As the MDSet theory has shown to be effective for the majority of linear and nonlinear systems [37, 50], it allows us to overlook dynamics process assumptions. Thus, dominating set theory is considered to be a more general network controllability method with better applicability.

In a symptom network constructed from multiple scales, a natural module structure corresponds to the original scales themselves. For instance, in our study that utilized four scales—PSS, GAD, PHQ, and ISI—can be thought of as 4 natural modules. However, there has functional overlaps between different items. For example, the PHQ scale, designed to measure levels of depression, includes items indirectly related to depression, such as sleep problems (PHQ3), which bears resemblance to the content assessed by the ISI scale. The essence of our module discovery method, therefore, lies in identifying modules formed at the data level rather than pre-assumed modules. Under this data-driven methodology, nodes with similar scoring trends are grouped into the same module. Meanwhile, existing clinical interventions and strategies do not treat each symptomatic item individually; rather, it target broader categories like depression, stress, anxiety, and sleep issues which are clustering by series of symptoms . This is in line with our module-based research philosophy - multiple symptoms will aggregate to form symptom clusters. Module control framework approach provides a higher-level perspective that allows for the analysis of symptom networks at a mesoscale—focusing not on individual nodes but on larger modules. While this might seem to offer a coarser granularity, it paradoxically yields clearer and more concise guidance for our interventions and treatments.

Upon analyzing the control capacity of the modules, it is evident that there is a control relationship from non-emotional modules to emotional modules within the psychopathology network. This is supported by the strong control phenomenon observed in both the *AMCS* and *ACF* dimensions of module control. These findings are consistent with prior research demonstrating a link from stress and sleep problems to anxiety and depression [51, 52]. Interventions aimed at non-emotional modules may effectively improve the state of the emotional modules, while the weak module control capacity of the emotional modules over non-emotional modules suggests that this process may be difficult to reverse. Three symptoms, namely ISI4, PSS2, and PSS7, were observed to have the highest *ACF* in non-emotional modules and can be considered as the primary controlling symptoms of the network. ISI4, which measures how satisfied/dissatisfied with their current sleep pattern. This means that the self-perception of sleep problems is the main controlling symptom in the sleep problem rather than a specific symptom (e.g., difficulty falling/staying asleep). As severe sleep problem can lead to emotional problem like depression and anxiety [51], psychological interventions aimed at reversing subjective feelings and perceptions of sleep problems may be more effective than pharmacological interventions [53].

In summary, we developed a new solution to study rampant comorbidity among discrete syndromes such as emotional and no-emotional problems. The results of this study can inform the development of more effective interventions for mental health in different disease population. It is noteworthy that the module control theory we propose is not exclusively for symptom networks but serves as a universal method and framework within the complex network domain. Beyond its application to symptom networks, we are committed to applying it to a broader array of real-world network systems.

**Limitations of the study**

Although module control offers potential benefits, it is crucial to acknowledge its limitations in real-world network applications. The applicability of our findings is, indeed, nuanced by the characteristics of our sample population, selection of scales and alongside our methodological assumption of similar MDSet weights. Varying combinations of inputs can result in different outcomes of the network construction, such as the emergence or dissolution of edges, or fluctuations in weights. These network variations inherently will influence the constitution of MDSets, hence the need for more effective network estimation methods that can fundamentally address the stability concerns in control theory. Meanwhile, module control methods should be further discussed in the context of other psychopathological approaches like phenomenological psychopathology [54, 55] and spatiotemporal psychopathology [56-58], to enhance its practical significance. The topological properties of networks can help us understand the connections between symptoms while revealing entities of underlying mental illness in the human body.In addition to symptom-level analysis, joint analysis with neuroimaging data will also be an important direction in the future.

In future endeavors, we aim to moving beyond cross-sectional analyses to explore longitudinal data, as this approach is more consistent with network theory's focus on within-person assumptions about disease etiology and treatment. Our current cross-sectional study lays the foundation for future work. Combined with our study of dynamic network control [59], the concept of module control can be expanded to understand how control power within a network changes and shifts over time, we can uncover novel insights into the disease's trajectory. This approach will not only enrich our comprehension of psychopathological mechanisms but also enhance the precision and effectiveness of psychosocial interventions.

## Acknowledgements

This work was supported by the National Natural Science Foundation of China [grant number 62176129]; the National Key Research and Development Program of the Ministry of Science and Technology of China [grant number 2022YFC2405603]; the National Science Fund for Distinguished Young Scholars [grant number 81725005]; the National Natural Science Foundation of China-Guangdong Joint Fund [grant number U20A6005]; and the Jiangsu Provincial Key Research and Development Program [grant number BE2021617].

## Author contribution

Xizhe Zhang conceived the study, drafted and revised the manuscript; Chunyu Pan, Quan Zhang and Yue Zhu performed data analysis, preparing the figures and drafted the manuscript; Shengzhou Kong revised the figures; Juan Liu performed data interpretation; Changsheng Zhang and Fei Wang review and editing the manuscript. All authors contributed substantially to the preparation of the manuscript.

## Conflict of interest statement

We declare that we have no financial and personal relationships with other people or organizations that can inappropriately influence our work, there is no professional or other personal interest of any nature or kind in any product, service and/or company that could be construed as influencing the position presented in, or the review of, the manuscript entitled, " *Module control of network analysis in psychopathology"*.

## Reference

[1] K. Kroenke, R. L. Spitzer, and J. B. W. Williams, "The PHQ-9: validity of a brief depression severity measure," *Journal of general internal medicine,* vol. 16, no. 9, pp. 606-613, 2001.

[2] M. B. First, J. B. W. Williams, R. S. Karg, and R. L. Spitzer, "User's guide for the structured clinical interview for DSM-5 disorders, research version (SCID-5-RV)," *Arlington: American Psychiatric Association,* 2015.

[3] R. J. McNally, *What is mental illness?* Harvard University Press, 2012.

[4] S. E. Hyman, "The diagnosis of mental disorders: the problem of reification," *Annual review of clinical psychology,* vol. 6, pp. 155-179, 2010.

[5] B. S. Held, "The distinction between psychological kinds and natural kinds revisited: can updated natural-kind theory help clinical psychological science and beyond meet psychology's philosophical challenges?," *Review of General Psychology,* vol. 21, no. 1, pp. 82-94, 2017.

[6] R. J. McNally, "Network analysis of psychopathology: Controversies and challenges," *Annual review of clinical psychology,* vol. 17, pp. 31-53, 2021.

[7] D. J. Robinaugh, R. H. A. Hoekstra, E. R. Toner, and D. Borsboom, "The network approach to psychopathology: a review of the literature 2008–2018 and an agenda for future research," *Psychological medicine,* vol. 50, no. 3, pp. 353-366, 2020.

[8] T. Cheung, Y. Jin, S. Lam, Z. Su, B. J. Hall, and Y.-T. Xiang, "Network analysis of depressive symptoms in Hong Kong residents during the COVID-19 pandemic," *Translational Psychiatry,* vol. 11, no. 1, p. 460, 2021.

[9] M. Belvederi Murri *et al.*, "Depressive symptom complexes of community-dwelling older adults: a latent network model," *Molecular Psychiatry,* vol. 27, no. 2, pp. 1075-1082, 2022.

[10] A. Rouquette *et al.*, "Emotional and behavioral symptom network structure in elementary school girls and association with anxiety disorders and depression in adolescence and early adulthood: a network analysis," *JAMA psychiatry,* vol. 75, no. 11, pp. 1173-1181, 2018.

[11] G. Briganti, C. Kornreich, and P. Linkowski, "A network structure of manic symptoms," *Brain and Behavior,* vol. 11, no. 3, p. e02010, 2021.

[12] L. Boschloo, R. A. Schoevers, C. D. van Borkulo, D. Borsboom, and A. J. J. J. o. A. P. Oldehinkel, "The network structure of psychopathology in a community sample of preadolescents," vol. 125, no. 4, p. 599, 2016.

[13] D. Borsboom, "A network theory of mental disorders," *World psychiatry,* vol. 16, no. 1, pp. 5-13, 2017.

[14] E. I. Fried, C. D. van Borkulo, A. O. J. Cramer, L. Boschloo, R. A. Schoevers, and D. Borsboom, "Mental disorders as networks of problems: a review of recent insights," *Social psychiatry and psychiatric epidemiology,* vol. 52, pp. 1-10, 2017.

[15] R. Quax, A. Apolloni, and P. M. Sloot, "The diminishing role of hubs in dynamical processes on complex networks," *Journal of The Royal Society Interface,* vol. 10, no. 88, p. 20130568, 2013.

[16] A. O. J. Cramer *et al.*, "Major Depression as a Complex Dynamic System," *PLOS ONE,* vol. 11, no. 12, p. e0167490, 2016, doi: 10.1371/journal.pone.0167490.

[17] D. Borsboom *et al.*, "Network analysis of multivariate data in psychological science," vol. 1, no. 1, p. 58, 2021.

[18] Y.-Y. Liu, J.-J. Slotine, and A.-L. Barabási, "Controllability of complex networks," *nature,* vol. 473, no. 7346, pp. 167-173, 2011.

[19] X. Zhang, T. Lv, X. Yang, and B. Zhang, "Structural controllability of complex networks based on preferential matching," *PloS one,* vol. 9, no. 11, p. e112039, 2014.

[20] X. Zhang, H. Wang, and T. Lv, "Efficient target control of complex networks based on preferential matching," *Plos one,* vol. 12, no. 4, p. e0175375, 2017.

[21] X. Zhang, J. Han, and W. Zhang, "An efficient algorithm for finding all possible input nodes for controlling complex networks," *Scientific reports,* vol. 7, no. 1, p. 10677, 2017.

[22] X. Zhang, T. Lv, and Y. Pu, "Input graph: the hidden geometry in controlling complex networks," *Scientific reports,* vol. 6, no. 1, p. 38209, 2016.

[23] C. Pan *et al.*, "Control analysis of protein-protein interaction network reveals potential regulatory targets for MYCN," *Frontiers in Oncology,* vol. 11, p. 633579, 2021.

[24] X. Zhang, "Altering indispensable proteins in controlling directed human protein interaction network," *IEEE/ACM Transactions on Computational Biology and Bioinformatics,* vol. 15, no. 6, pp. 2074-2078, 2018.

[25] W.-F. Guo, X. Yu, Q.-Q. Shi, J. Liang, S.-W. Zhang, and T. Zeng, "Performance assessment of sample-specific network control methods for bulk and single-cell biological data analysis," *PLoS Computational Biology,* vol. 17, no. 5, p. e1008962, 2021.

[26] Q. Li *et al.*, "Controllability of Functional Brain Networks and Its Clinical Significance in First-Episode Schizophrenia," *Schizophrenia Bulletin,* 2022.

[27] R. Tang *et al.*, "Longitudinal association of executive function and structural network controllability in the aging brain," *GeroScience,* pp. 1-13, 2022.

[28] B. Lee, U. Kang, H. Chang, and K.-H. Cho, "The hidden control architecture of complex brain networks," *Iscience,* vol. 13, pp. 154-162, 2019.

[29] C. Pan, Y. Ma, L. Wang, Y. Zhang, F. Wang, and X. Zhang, "From Connectivity to Controllability: Unraveling the Brain Biomarkers of Major Depressive Disorder," *Brain Sciences,* vol. 14, no. 5, p. 509, 2024.

[30] Y.-Y. Ding *et al.*, "Network analysis reveals synergistic genetic dependencies for rational combination therapy in Philadelphia chromosome-like acute lymphoblastic leukemia," *Clinical cancer research: an official journal of the American Association for Cancer Research,* vol. 27, no. 18, p. 5109, 2021.

[31] W.-F. Guo, S.-W. Zhang, T. Zeng, T. Akutsu, and L. Chen, "Network control principles for identifying personalized driver genes in cancer," *Briefings in bioinformatics,* vol. 21, no. 5, pp. 1641-1662, 2020.

[32] X. Zhang *et al.*, "Cancer-keeper genes as therapeutic targets," *Iscience,* vol. 26, no. 8, 2023.

[33] F. Fang, B. Godlewska, R. Y. Cho, S. I. Savitz, S. Selvaraj, and Y. Zhang, "Personalizing repetitive transcranial magnetic stimulation for precision depression treatment based on functional brain network controllability and optimal control analysis," *Neuroimage,* vol. 260, p. 119465, 2022.

[34] V.-B. Popescu, K. Kanhaiya, D. I. Năstac, E. Czeizler, and I. Petre, "Network controllability solutions for computational drug repurposing using genetic algorithms," *Scientific reports,* vol. 12, no. 1, p. 1437, 2022.

[35] X. Wei, C. Pan, X. Zhang, and W. Zhang, "Total network controllability analysis discovers explainable drugs for Covid-19 treatment," *Biology Direct,* vol. 18, no. 1, pp. 1-14, 2023.

[36] M. E. J. Newman, "Modularity and community structure in networks," *Proceedings of the national academy of sciences,* vol. 103, no. 23, pp. 8577-8582, 2006.

[37] J. C. Nacher and T. Akutsu, "Dominating scale-free networks with variable scaling exponent: heterogeneous networks are not difficult to control," *New Journal of Physics,* vol. 14, no. 7, p. 073005, 2012/07/03 2012, doi: 10.1088/1367-2630/14/7/073005.

[38] L. Guan and H. Wang, "A heuristic approximation algorithm of minimum dominating set based on rough set theory," *Journal of Combinatorial Optimization,* vol. 44, no. 1, pp. 752-769, 2022.

[39] N. Williams, "The GAD-7 questionnaire," *Occupational medicine,* vol. 64, no. 3, pp. 224-224, 2014.

[40] D. S. F. Yu, "Insomnia Severity Index: psychometric properties with Chinese community-dwelling older people," (in English), *JOURNAL OF ADVANCED NURSING,* vol. 66, no. 10, pp. 2350-2359, OCT 2010, doi: 10.1111/j.1365-2648.2010.05394.x.

[41] R. Y. M. Cheung, "Patient Health Questionnaire-9 (PHQ-9)," in *Handbook of Assessment in Mindfulness Research*, O. N. Medvedev, C. U. Krägeloh, R. J. Siegert, and N. N. Singh Eds. Cham: Springer International Publishing, 2022, pp. 1-11.

[42] S. Townsend and O. N. Medvedev, "Perceived Stress Scale (PSS)," in *Handbook of Assessment in Mindfulness Research*, O. N. Medvedev, C. U. Krägeloh, R. J. Siegert, and N. N. Singh Eds. Cham: Springer International Publishing, 2022, pp. 1-13.

[43] J. Friedman, T. Hastie, and R. Tibshirani, "Sparse inverse covariance estimation with the graphical lasso," (in English), *BIOSTATISTICS,* vol. 9, no. 3, pp. 432-441, JUL 2008, doi: 10.1093/biostatistics/kxm045.

[44] R. Lambiotte and M. T. Schaub, *Modularity and Dynamics on Complex Networks*, Cambridge: Cambridge University Press, 2022. [Online]. Available: https://www.cambridge.org/core/elements/modularity-and-dynamics-on-complex-networks/2CD4E8743B0477CBE9157E1E0A4DC926.

[45] A. S. d. Mata, "Complex Networks: a Mini-review," *Brazilian Journal of Physics,* vol. 50, no. 5, pp. 658-672, 2020/10/01 2020, doi: 10.1007/s13538-020-00772-9.

[46] S. Epskamp, D. Borsboom, and E. I. Fried, "Estimating psychological networks and their accuracy: A tutorial paper," (in English), *BEHAVIOR RESEARCH METHODS,* vol. 50, no. 1, pp. 195-212, FEB 2018, doi: 10.3758/s13428-017-0862-1.

[47] L. F. Bringmann *et al.*, "What do centrality measures measure in psychological networks?," *Journal of abnormal psychology,* vol. 128, no. 8, p. 892, 2019.

[48] S. K. Grady, F. N. Abu-Khzam, R. D. Hagan, H. Shams, and M. A. Langston, "Domination based classification algorithms for the controllability analysis of biological interaction networks," *Scientific Reports,* vol. 12, no. 1, pp. 1-10, 2022.

[49] S. Bakhteh, A. Ghaffari-Hadigheh, and N. Chaparzadeh, "Identification of minimum set of master regulatory genes in gene regulatory networks," *IEEE/ACM Transactions on Computational Biology and Bioinformatics,* vol. 17, no. 3, pp. 999-1009, 2018.

[50] V. Ravindran *et al.*, "Network controllability analysis of intracellular signalling reveals viruses are actively controlling molecular systems," *Scientific reports,* vol. 9, no. 1, pp. 1-11, 2019.

[51] A. J. Scott, T. L. Webb, M. Martyn-St James, G. Rowse, and S. Weich, "Improving sleep quality leads to better mental health: A meta-analysis of randomised controlled trials," *Sleep Medicine Reviews,* vol. 60, p. 101556, 2021.

[52] L. J. S. Schweren *et al.*, "Diet quality, stress and common mental health problems: A cohort study of

[53] 121,008 adults," *Clinical Nutrition,* vol. 40, no. 3, pp. 901-906, 2021.

[53] Y. Zhang *et al.*, "Comparative efficacy and acceptability of psychotherapies, pharmacotherapies, and their combination for the treatment of adult insomnia: A systematic review and network meta-analysis," *Sleep Medicine Reviews,* p. 101687, 2022.

[54] G. Stanghellini, M. Broome, A. Raballo, A. V. Fernandez, P. Fusar-Poli, and R. Rosfort, *The Oxford handbook of phenomenological psychopathology*. Oxford University Press, USA, 2019.

[55] G. Messas, M. Tamelini, M. Mancini, and G. Stanghellini, "New Perspectives in Phenomenological Psychopathology: Its Use in Psychiatric Treatment," (in English), *Frontiers in Psychiatry,* Perspective vol. 9, 2018-September-28 2018, doi: 10.3389/fpsyt.2018.00466.

[56] G. Northoff, J. Daub, and D. Hirjak, "Overcoming the translational crisis of contemporary psychiatry – converging phenomenological and spatiotemporal psychopathology," *Molecular Psychiatry,* vol. 28, no. 11, pp. 4492-4499, 2023/11/01 2023, doi: 10.1038/s41380-023-02245-2.

[57] G. Northoff, "Spatiotemporal Psychopathology–A Novel Approach to Brain and Symptoms," *Archives of Neuropsychiatry,* vol. 59, no. Suppl 1, p. S3, 2022.

[58] G. Northoff and D. Hirjak, "Spatiotemporal Psychopathology–An integrated brain-mind approach and catatonia," *Schizophrenia research,* 2022.

[59] C. Pan, Z. Su, H. Zheng, C. Zhang, W. Zhang, and X. Zhang, "Adaptive control of dynamic networks," *arXiv preprint arXiv:2302.09743,* 2024.

[60] F. Pedregosa *et al.*, "Scikit-learn: Machine Learning in Python," (in English), *JOURNAL OF MACHINE LEARNING RESEARCH,* vol. 12, pp. 2825-2830, OCT 2011.

[61] V. D. Blondel, J. L. Guillaume, R. Lambiotte, and E. Lefebvre, "Fast unfolding of communities in large networks," (in English), *JOURNAL OF STATISTICAL MECHANICS-THEORY AND EXPERIMENT,* OCT 2008, Art no. P10008, doi: 10.1088/1742-5468/2008/10/P10008.

[62] R. E. Kalman, "Mathematical description of linear dynamical systems," *Journal of the Society for Industrial and Applied Mathematics, Series A: Control,* vol. 1, no. 2, pp. 152-192, 1963.

[63] M. R. Chernick, *Bootstrap methods: A guide for practitioners and researchers*. John Wiley & Sons, 2011.

[64] S. P. Borgatti and D. S. Halgin, "Analyzing Affiliation Networks," 2011.

[65] U. Brandes, "On variants of shortest-path betweenness centrality and their generic computation," (in English), *SOCIAL NETWORKS,* vol. 30, no. 2, pp. 136-145, MAY 2008, doi: 10.1016/j.socnet.2007.11.001.

[66] L. C. Freeman, "Centrality in social networks conceptual clarification," *Social Networks,* vol. 1, no. 3, pp. 215-239, 1978/01/01/ 1978, doi: https://doi.org/10.1016/0378-8733(78)90021-7.

[67] J. Saramaki, M. Kivela, J. P. Onnela, K. Kaski, and J. Kertesz, "Generalizations of the clustering coefficient to weighted complex networks," (in English), *PHYSICAL REVIEW E,* vol. 75, no. 2, FEB 2007, Art no. 027105, doi: 10.1103/PhysRevE.75.027105.

[68] V. Batagelj and M. Zaversnik, "Fast algorithms for determining (generalized) core groups in social networks," (in English), *ADVANCES IN DATA ANALYSIS AND CLASSIFICATION,* vol. 5, no. 2, pp. 129-145, JUL 2011, doi: 10.1007/s11634-010-0079-y.

[69] A. Hagberg, P. Swart, and D. S Chult, "Exploring network structure, dynamics, and function using NetworkX," Los Alamos National Lab.(LANL), Los Alamos, NM (United States), 2008.

## STAR ★ Methods

### Key Resources Table

| REAGENT or RESOURCE | SOURCE | IDENTIFIER |
| --- | --- | --- |
| Biological samples | | |
| Participant scale score | Study participants | N/A |
| Software and algorithms | | |
| Python version 3.7.5 | Python Software Foundation | https://www.python.org/ |

| Scikit-learn version 1.0.2 | Scikit-learn Developers (open-source) | https://scikit-learn.org/stable/ |
| Visio version 2013 | Microsoft | https://www.microsoft.com/en-us/microsoft-365/visio/flowchart-software |
| Gephi version 0.9.2 | Scikit-learn Developers (open-source) | https://gephi.org/ |
| Origin version 2021 | Origin Lab | https://www.originlab.com/ |

## Resource Availability

### Lead contact

Further information and requests for resources should be directed to and will be fulfilled by the lead contact, Xizhe Zhang (zhangxizhe@njmu.edu.cn).

### Materials availability

The study did not generate new unique reagents.

### Data and code availability

The software for this study are freely available in the GitHub public repository at https://github.com/network-control-lab/module-control.

## Experimental Model and Study Participant Details

### Participants

A cross-sectional survey was conducted on first-year medical students at Nanjing Medical University from September 24th to September 28th, 2022. The study utilized an online questionnaire which was administered through the WeChat official account platform. College students were notified by their school counselor-teachers that they would receive a link to the questionnaire program electronically and were asked to complete these questionnaires within a time limit. Prior to completing the questionnaire, participants provided online informed consent and the study was approved by the Biomedical Ethics Committee of Nanjing Medical University (2022793). Despite the short duration of data collection, a total of 2773 valid questionnaires were collected for analysis.

### Measures

The study employed various self-report questionnaires to measure the presence and severity of depressive, anxiety, insomnia, and stress symptoms among the participants. The Patient Health Questionnaire 9 (PHQ-9) was utilized to assess depressive symptoms in the last two weeks, with a total score ranging from 0 to 27 [41]. Participants with a total score below 5 were considered as without depressive symptoms, while those with scores between 5-9, 10-14, and greater than 14 were identified as with mild, moderate, and major depression, respectively. The Generalized Anxiety Disorder-7 (GAD-7) was employed to evaluate anxiety symptoms in the last two weeks, with a total score range between 0 and 27 [39]. Participants with a total score below 5 were identified as without anxiety symptoms, while those with scores between 5-9, 10-13, and greater than 13 were classified as with mild, moderate, and severe anxiety, respectively. Furthermore, the Insomnia Severity Index (ISI) was utilized to measure the severity of insomnia in the last two weeks, with a total score ranging from 0 to 28. Participants with a total score of less than 8, 8-14, 15-21, and greater than 21 were defined as having no, mild, moderate, and severe insomnia, respectively [40]. Finally, the Perceived Stress Scale 14 (PSS-14) was used to assess stress levels in the past month. The PSS-14 consists of 14 items on a 5-point scale, with a total score range of 0 to 56. Participants with a total score below 29 were classified as having a normal level of perceived stress, while those scoring between 29-42 and greater than 42 were identified as having moderate and severe levels of stress perception, respectively [42].

In this study, we selected these scales for the following reasons: Firstly the data used in the study were from a large epidemiological survey of mood symptoms in Chinese university students during the epidemic, the primary aim of this survey was to get a quick snapshot of the mood and sleep status of university students in the context of the epidemic lockdown, so we used short screening scales; and secondly we chose the scales on the basis of their relevance to psychopathological symptoms of the mood screen for depression and anxiety as well as the most at-risk suicidal behaviors, and their widespread use in research and clinical settings.

**Methods Details**

*Network estimation*

We employed the Graphical LASSO method [43], a common approach for constructing symptom networks, to create a network based on individual item scores and their associations [17]. The Graphical LASSO method estimates a Gaussian Graphical Model and applies regularization through the use of the Least Absolute Shrinkage and Selection Operator (LASSO) and results in a sparse inverse covariance matrix. The LASSO penalty (L1 penalty) helps to eliminate potential bias and create a sparse network, which has been widely used in psychological research [17]. The mathematical formulation for this method is as follows [43]:

$$\widehat{K} = argmin_K(trSK - \log det K + \alpha \|K\|_1)$$

Where $K$ is the sparse inverse covariance matrix to be estimated; $S$ is the covariance matrix of the sample we used; $\|K\|_1$ is the sum of the absolute values of off-diagonal coefficients of $K$; $\alpha$ is the sparsity parameter where we set $\alpha$=0.05. The Graphical Lasso estimator is available in Scikit-learn [60] and coded in Python.

In order to improve the accuracy of analysis, we absolutized the partial correlation coefficients in the inverse covariance matrix and set the matrix diagonals to zero. This approach allows us to focus solely on the absolute association strength of the edges and ignore negative or positive associations. Additionally, we only examined the strength of the association between different nodes, disregarding the association with itself. Based on the preprocessed sparse inverse covariance matrix, we constructed the network. For any node pair *(i, j)*, if the corresponding element in the matrix was not zero, an edge was considered to exist between *(i, j)* and the weight of the edge was the value of the corresponding element in the matrix.

*Module detection*

The Louvain algorithm [61] was adopted to reclassify the items into different modules, it is a disjoint module detection method based on modularity. For a network, the modularity of its module represents the difference between the number of edges within the module and the number of edges in a random case. The formula as follows [61]:

$$Q = \frac{1}{2m}\sum_{ij}\left[A_{ij} - \frac{k_i - k_j}{2m}\right]\delta(c_i, c_j)$$

where $A_{ij}$ represents the weight of the edge between *i* and *j*, $k_i = \sum_j A_{ij}$ is the sum of the weights of the edges belongs to node *i*, $c_i$ is the community to which node *i* is assigned, $\delta(c_i, c_j)$ is 1 if $c_i = c_j$ and 0 otherwise, $m = \frac{1}{2}\sum_j A_{ij}$ represents the total weight of the network.

The Louvain algorithm maximizes the modularity $Q$ by using a greedy strategy to find the best partition through iterations. Specifically, the algorithm assigns a different module to each node of the network. By calculating the gain in modularity $\Delta Q$, the algorithm selects a neighboring module with the largest $\Delta Q > 0$ to join. This process is repeated until the modularity $Q$ no longer changes. The gain in modularity $\Delta Q$ can be represented mathematically as follows:

$$\Delta Q = \left[\frac{\sum in + k_{i,in}}{2m} - \left(\frac{\sum tot + k_i}{2m}\right)^2\right] - \left[\frac{\sum in}{2m} - \left(\frac{\sum tot}{2m}\right)^2 - \left(\frac{k_i}{2m}\right)^2\right] = \left[\frac{k_{i,in}}{2m} - \frac{\sum tot\, k_i}{2m^2}\right]$$

where the $\sum in$ represents the sum of the weights of the edges inside module *c*, $\sum tot$ represents the sum of the weights of the nodes in module *c*, $k_i$ is the weights of the node *i*, $k_{i,in}$ represents the sum of the weight of the edges from *i* to nodes in module *c* and *m* is the total weight of the network. The algorithm subsequently merges all nodes that belong to the same module into a new node and repeats the first step until the modularity $Q$ no longer changes. This process continues until the final module assignment for each node is obtained.

*Controllability of symptom network*

A network is controllable if it can be driven from any initial state to the desired state by inputting external control signals [62] and the nodes used to receive external signals are called driver nodes (Figure 1.A). For a directed network with linear dynamic, the minimum driver nodes set can be found through structural controllability to achieve control it [18]. For undirected networks, the dominating set theory can be used to find the fewest driver nodes in the network [48]. In this model, each driver node can control its links individually and can control neighboring nodes and itself (Figure 1.B and 1.D). Since the dominating set theory has proven effective for the majority of linear and nonlinear systems [37, 50], it is possible to disregard the influence of dynamics on network controllability and solely focus on the structural control of the network. For the undirected symptom network we use, each edge can be considered as bi-directional and the network is structurally controllable by selecting the nodes in an minimum dominating set (MDSet) as driver nodes [37].

*Minimum dominating set*

For an undirected network $G = (V, E)$, a subset of nodes $D \subseteq V$ is defined as the dominating set if every node $n \in V$ is belongs to subset $D$ or connected to any node which belongs to subset $D$. In other words, the union set of the controlling areas of the dominating set will cover all the nodes in the network (Figure 1.C). By selecting the node in the dominating set as the driver node, the entire network can be driven [37]. The smallest size of dominating set of the network is called minimum dominating set (MDSet) whose control is sufficient to fully control the system's dynamics with the lowest control cost (Figure 1.C).

We employed a brute-force search method to discover all MDSets. In the brute-force search processing, we began by searching for the smallest possible MDSet size, i.e., a single node $N_1 = \{n_1\}$, and checking if it can cover all nodes in the network. If it can, the MDSet size is 1, and we proceed to try every single node $N_1$ to determine all MDSets. If any $N_1$ fails to cover all nodes in the network, we attempt combinations of two nodes $N_2 = \{n_1, n_2\}$. If any $N_2$ can cover all nodes in the network, the MDSet size is 2, and we try every possible $N_2$ to determine all MDSets. If any $N_2$ fails to cover all nodes in the network, we repeat the process until we find a node set $N_m$ that can serve as the MDSet for the network, with $m$ being the minimum size of the dominating nodes set. For each MDSet, its weight is defined as the sum of the weights of each dominating node in this MDSet.

*Module Control Network and Control Frequency*

Consider an undirected network $G = (V, E)$ with $M$ modules, for a given MDSet $MDS$, the driver nodes in $MDS$ normally lies in different modules (Figure 2.C). The control areas of any driver node n represent as $A_n$, which is a node set include neighbors of $n$ and itself (Figure 1.B). For a module $m$, we denote the driver nodes within the module as $MDS_m = MDS \cap m$, the control areas of module $m$ can be represented as $AC_m = \cup_{n \in MDS_m} A_n$, which represent the union control areas of driver nodes in module $m$ (Figure 2.A).

The control areas $AC_m$ of module $m$ often expands to the nodes of other modules (Figure 2.A). To analysis the control power from one module to another, we present a novel metric, Module Control Strength (MCS). This metric quantifies the percentage of controlled nodes in module $m'$ that fall within the control areas $AC_m$ of module $m$ (Figure 2.B). For a given MDSet MDS, the $MCS_{m,m'}$ from module m to module $m'$ can be represented by:

$$MCS^{MDS}_{m,m'} = \frac{|AC_m \cap V_{m'}|}{|V_{m'}|}$$

Since there are numerous different MDSets may exist in the network (Figure 2.D), the Average Module Control Strength (AMCS) is introduced to account for all MDSets within a network:

$$AMCS_{m,m'} = \frac{\sum_{MDS \in \mathfrak{D}} MCS^{MDS}_{m,m'}}{|\mathfrak{D}|} \in [0,1]$$

where $\mathfrak{D}$ denote the set of all MDSets of a network. The AMCS measures the mean control strength from module m to m' under all MDSets of a network.

Based on the concept of AMCS, we have developed a novel module scale network, which we refer to as the **M**odule **C**ontrol **N**etwork (MCN, Figure 2.E). The nodes of the MCN correspond to the modules in the original network, while the edges weights depict the AMCS that exist between these modules (Figure 2.E). It is important to highlight that the control strength between two modules in a given MDSet, represented as $MCS_{m,m'}$ and $MCS_{m',m}$, is not symmetrical (Figure 2.C). This asymmetry stems from the fact that the number of driver nodes within each module may vary. Consequently, the MCN can be characterized as a directed network, allowing for a more comprehensive understanding of the control dynamics and relationships between modules in the network under investigation.

We also define the control frequency of each node within the module, which is considered as indicator of the node importance. The control frequency of node n can be represented as

$$CF_n = \frac{1}{|\mathfrak{D}|} \sum_{i=1}^{|\mathfrak{D}|} \sum_{j=1}^{|D_i|} [n \in D_i] \cdot [n = D_{i,j}]$$

where $[n \in D_i]$ is the indicator function, which equals 1 if node *n* is present in $D_i$ and 0 otherwise. $[n = D_{i,j}]$ is another indicator function, which equals 1 if element *n* is equal to the *j*-th element in set $D_i$ and 0 otherwise. Based on above definition, we can also compute the control frequency of each module based on the driver nodes within it. Therefore, we propose the *Average module Control Frequency* (*ACF*) as another indicator. For module $C_i$ with node set $V_{C_i}$, it can be represented as

$$ACF_i = \frac{\sum_{n \in V_{C_i}} CF_n}{|V_{C_i}|}$$

By utilizing *MCN* and *ACF* we can assess the control capacity of each module, we applied it to a symptom network built from real-world data and utilized this approach to identify the main controlling symptoms and items within the network.

*Bootstrapping*

For the non-parametric bootstrapping [63], we randomly selected 80% of the full sample and re-estimated the network to calculate metrics, including edge weights, MDSets size, and module numbers, etc. This process was repeated 1,000 times to obtain bootstrapping results such as distribution, 95% confidence intervals, and mean values for each metric.

The case-dropping subset bootstrapping [46] calculated the average correlation (also known as correlation stability coefficient, CS-C) between the metrics of the full sample network and the metrics of 1,000 bootstrapped networks. The sample size for constructing the bootstrapped network was gradually reduced to observe the stability of CS-C (sampled size: 50%-90%, step: 10%). If the metrics of the network did not change significantly after excluding part of the sample, the metrics could be considered stable. The CS-C represented the maximum proportion of samples that could be removed, such that with 95% probability, the correlation between original centrality indices could reach at least 0.25 is generally preferred, with a value above 0.5 being optimal [46].

**Quantification and Statistical Analysis**

The degree centrality [64] of node *i* measures the proportion of node that linked to all other nodes and it is determined by the number of neighbors connected to the node, represented as $DC_i = \frac{degree_i}{n-1}$, where *n* is the number of nodes in the network. The average strength of node i can be represented as $AS_i = \frac{\sum A_i}{DC_i}$, where $\sum A_i$ represents the total strength of node *i*.

Betweenness centrality [65] and closeness centrality [66] are both related to network shortest paths. The betweenness centrality of node *i* is calculated as $BC_i = \sum_{s \neq i \neq t \in n} \frac{p_{s,t}(i)}{p_{s,t}}$, where $p_{s,t}(i)$ is the number of shortest paths between nodes s and t going through node *i*, and $p_{s,t}$ is the number of all shortest paths between nodes *s* and *t*. The closeness centrality of node *i* is represented as $CC_i = \frac{r-1}{n-1}\frac{r-1}{\sum d_{i,j}}$, where $d_{i,j}$ is the length of the shortest path between nodes *i* and *j*, and *r* is the number of nodes reachable from node *i*.

Clustering coefficient [67] and K-core [68] are related to network structures. The Clustering Coefficient is calculated as $CL_i = \frac{2T_i}{d_i(d_i-1)}$, where $T_i$ is the number of triangles including node *i* and $d_i$ is the degree of node *i*. The K-core of a node corresponds to the largest subnet with a node degree of *K* or greater. The core value of a node is the largest value *K* containing the node. All these indicators are provided by the networkx [69] and coded in Python.

In order to enhance the comparability of different network indices, we use z-score to standardize the above six indices and z-score can be expressed as follows:

$$\text{z} - \text{score} = \frac{x - \bar{X}}{SD}$$

where $x$ represents the sample value to be normalised; $\bar{X}$ represents the mean value of all samples; $SD$ represents the standard deviation of all sample values.